# Some polarized lines of the second solar spectrum (SrI, CaI, BaII, $C_2$, MgH, NdII) observed at the Meudon Solar Tower spectropolarimeter

*Jean-Marie Malherbe (emeritus astronomer)*

Observatoire de Paris, PSL Research University, LIRA, France

Email: Jean-Marie.Malherbe@obspm.fr; ORCID: https://orcid.org/0000-0002-4180-3729

9 June 2026

**ABSTRACT**

The second solar spectrum is the spectrum of the Stokes parameter Q (linear polarization) close to the solar limb. It is made of a few polarized lines with Q/I of about 1% (such as CaI, SrI, or BaII), but most lines exhibit weaker polarization. This paper presents processing of unpublished observations made in 2008 with the Meudon solar tower spectropolarimeter, which are of interest for weak and turbulent unresolved magnetic field measurements in the quiet Sun, through the Hanle effect. All data shown in this paper are available in the on-line dataset for further investigations at:

https://entrepot.recherche.data.gouv.fr/dataset.xhtml?persistentId=doi:10.57745/YK18JE

## I - INTRODUCTION

Scattering processes on the Sun are sources of polarization in the solar spectrum (Stenflo & Keller, 1997; Stenflo, 2004). The polarization is zero at disk centre (symmetry of the radiation field) and increases towards the solar limb (where the line of sight is orthogonal to the incident radiation field of the photosphere). The Stokes parameter Q represents linear polarization parallel to the limb, and the ratio Q/I is the polarization rate. To determine this polarization, the slit of our solar spectrograph was parallel to the limb, at various distances (5 or 20 arcsec), and the camera integrated for a long time to reach sufficient signal to noise ratio (SNR). After passing through the polarimeter, the beam was injected into the large 14 m spectrograph. The dispersive element was an echelle grating which offers about R = 350000 spectral resolution (or 8 mA) in the blue part of the spectrum. Our spectral sampling was 5-7 mA, and the spatial sampling along the slit 0.5". The observed lines are of interest for the measurements of weak (tens of Gauss) and turbulent (unresolved) magnetic fields of the quiet Sun owing to the Hanle effect (Stenflo, 1982). In the presence of magnetic fields, one observes a depolarization and a rotation of the polarization plane; as our observations do not include Stokes U, they are only useful for the depolarization of the Q parameter. We observed CaI 4227 A, BaII 4554 A, SrI 4607 A, $C_2$ and MgH molecules at 5140 A and NdI 5250 A. Results are presented after a short description of the instrumental procedure.

## II - OBSERVATIONS WITH THE SPECTROPOLARIMETER

The Meudon solar tower includes a coelostat (2 flat mirrors of 80 cm diameter) which feed a 60 cm aperture vertical telescope of f = 45 m focal length (figure 1). However, it is used most time with a focal reducer 2 (f = 22.5 m) more adapted to the Meudon seeing (in that case the telescope is diaphragmed at 30 cm, so that the beam aperture is always f/75 in the image plane). The primary image of the Sun (F1 focus, 42 cm or 21 cm image diameter with the focal reducer) forms after two reflections on a Newton mirror under normal incidence and a second flat mirror under 45° incidence.

The spectrograph (figure 2) has a focal length of 14 m; this large value was chosen many decades ago for photographic films (70 mm). Today with CCD or CMOS sensors it is not useful, so that there is a reduction of the spectrum size at the output (0.16 magnification). The grating has 300 grooves/mm and a blaze angle of 63°26' working in orders 5 to 15 (the orders are selected by narrow bandpass filters). In the blue part of the spectrum ($\lambda < 450$ nm), the dispersion is larger than 12 mm/A (before spectrum reduction). The usual slit has a width of 0.15 mm (0.7" or 1.4" on the Sun), allowing to reach the resolving power R = 350000. This slit width provides the spectral resolution of 8 mA or better in the blue. We used a cooled CCD sensor of Princeton Instruments, with mechanical shutter, based on the Kodak KAF1600 chip (85000 electrons full well capacity, pixels of 9 x 9 microns$^2$, 12 bits digitization). With this sensor, the spectral sampling is 4.6 mA/pixel at $\lambda = 450$ nm. In the spatial direction along the slit, the spectral sampling is 0.25"/pixel (or 0.5"/pixel with the focal reducer of the telescope, this is the case of this work).

The polarimeter (figure 3) is located in the primary image at f/75. It is made of one Liquid Crystal Variable Retarder (LCVR, 40 mm diameter) and a precision dichroic linear polarizer from Meadowlark company (single beam polarimetry, Malherbe *et al*, 2007). Then the beam is injected into the spectrograph. The polarimeter can run at 50 Hz but is limited by the speed of the camera (1 Hz) and essentially by the exposure time which is rather long in the blue (10 s typical) because of bad mirror coatings (for instance, 20 s at 420 nm are necessary for SNR = 170 in the continuum, or SNR = 70 in the CaI 4227 A line core; and observations of CaII H and K are impossible in the violet). The polarimeter has no chromatism, because the voltage (figure 4) is adapted to the observed line to produce exactly zero, quarter or half wave retardance.

The polarimeter delivered the combinations I+Q and I-Q in sequence. The driving voltages were chosen for each line so that the LCVR was strictly modulating between zero and half wave. Several tens of observations obtained within a loop were summed and then a final integration was made along the slit in order to reach a final SNR better than 10000. For instance, with 36 accumulations of spectra ($\lambda$, x) and 300 pixels summation (along the slit in x direction), we got in the core of CaI 4227 A line SNR = 7000.

Observations were performed with the slit parallel to the limb. Dark currents were subtracted and Flat Fields (FF) were measured at disk centre. FF are useful to correct the transmission between the two states of the LCVR delivering alternatively I+Q and I-Q. Flat Fields were observed systematically at disk centre in the quiet Sun (where the linear polarization should be zero) after each limb sequence. The FF was systematically applied when available.

**Limitations of our observations**

Our CCD camera was slow, and the exposure times in the blue were long, so that the seeing may alter the spectra and a small spatial shift may occur between I+Q and I-Q measurements.

The coelostat may be the source of variable and parasitic polarization because of the evolving incidence on the two mirrors during the diurnal motion of the Sun. Hence, crosstalk (mainly between Stokes Q and U) may occur. However our calculations showed us that it vanishes at the meridian (at about 12:00 UT for Meudon) for any season, or at the equinox (0° declination) for any time. Most observations were performed in May 2008 (18° solar declination, so that instrumental polarization vanished only at the meridian); figures 5 & 6 report what happens (theoretically) during the day for an incident beam of blue light at 460 nm, linearly polarized with either Q = 0 or U = 0. Of course, we do not know the exact polarization state of the incident light on the coelostat. As we measure only Q and I, we may lose some information about the linear polarization rate $(Q^2+U^2)^{1/2}/I$. Our calculations involve the knowledge of the complex refractive index of Aluminium coatings: we took a theoretical value, as it is impossible to know the exact state of the mirrors since coatings are old and deteriorate progressively with time, especially outside.

Also, the telescope has no guiding system, so that the limb distance of the slit is known only at the beginning of the run; there is a small drift during the run which is manually compensated, so that distance fluctuations occur during observations.

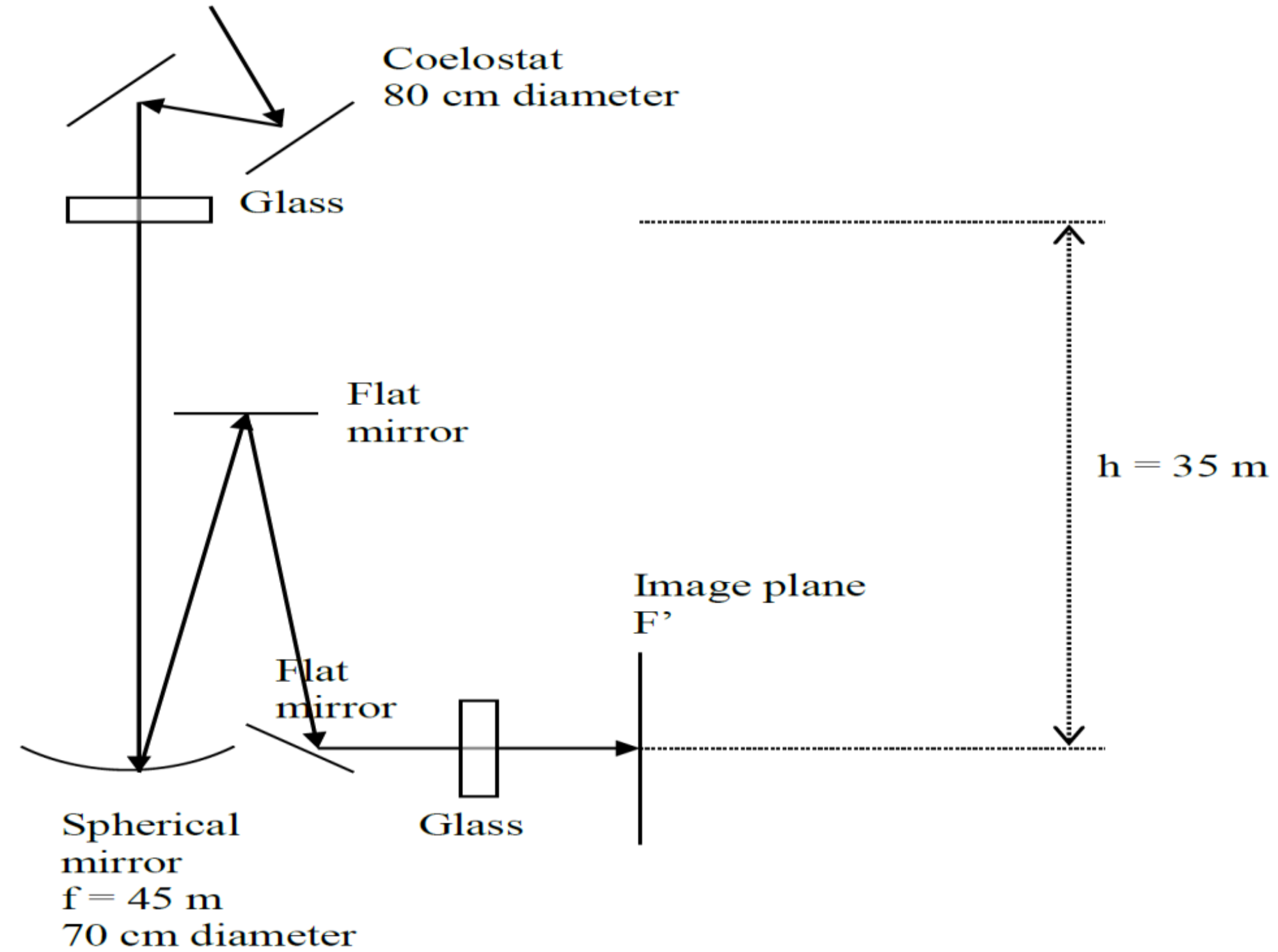

*Figure 1 : Meudon solar tower and its telescope*

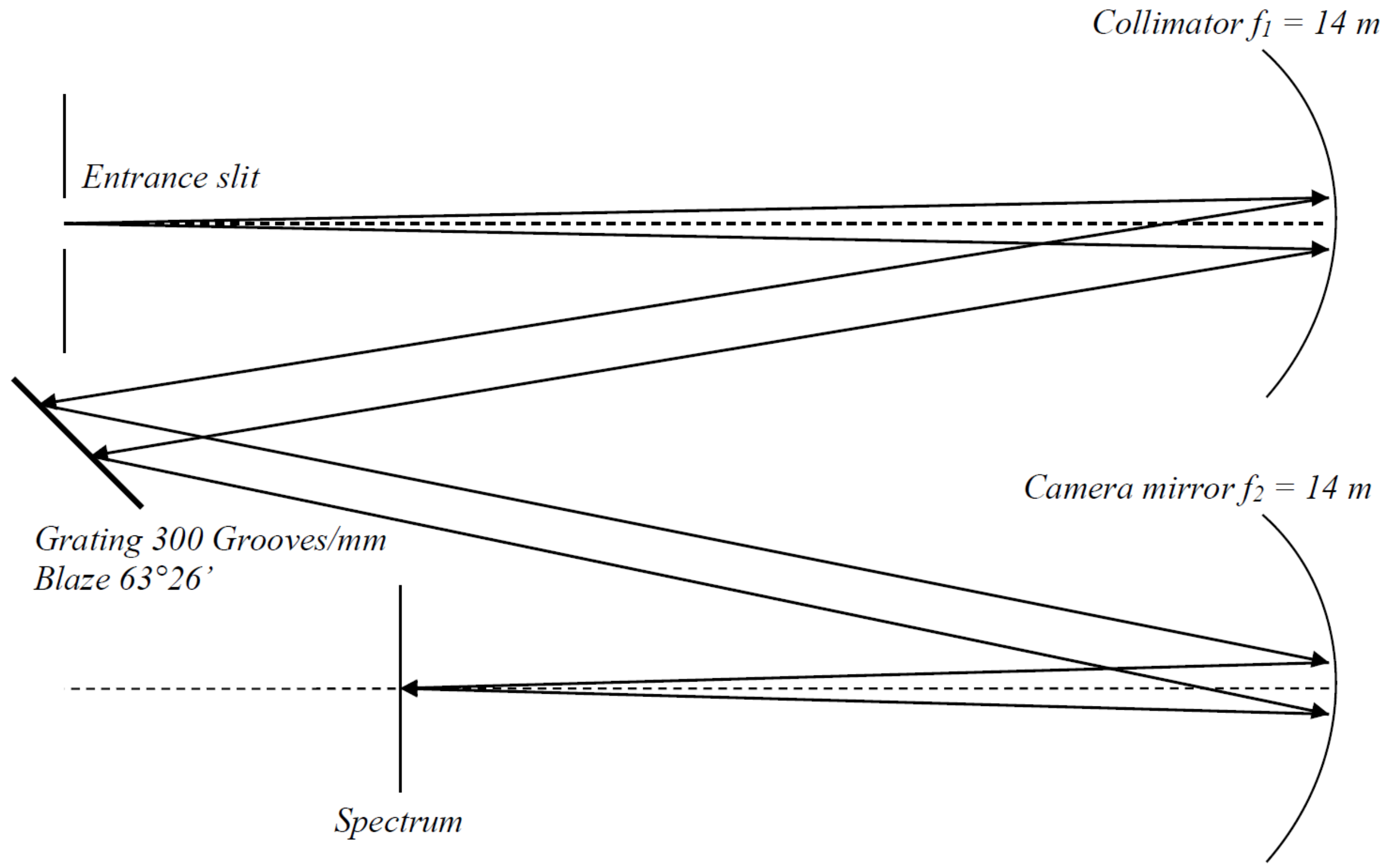

*Figure 2 : The 14 m horizontal spectrograph of Meudon solar tower*

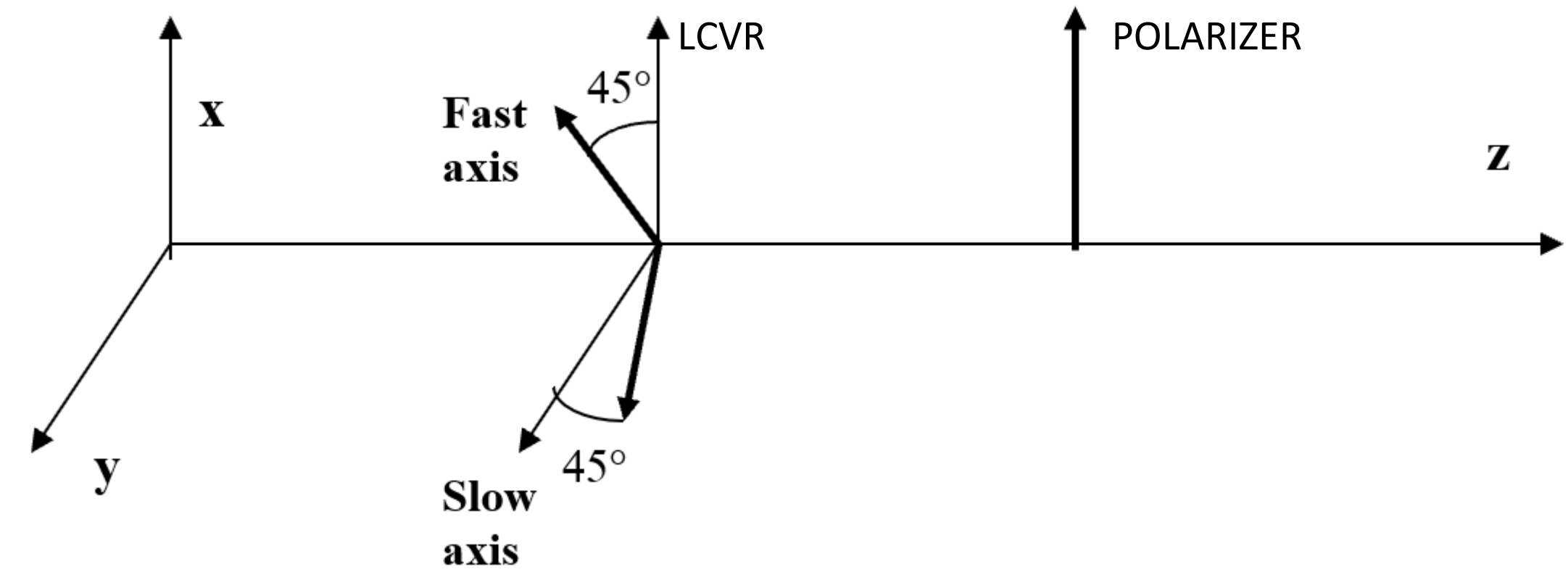

*Figure 3 : The single beam polarimeter of Meudon solar tower located in the image plane F' of Figure 1 before injection into the spectrograph. The LCVR from Meadowlark has a clear aperture of 40 mm. It can modulate between two chosen states in the range 0 to 1 wave, depending of the voltage (0 to 10 V) applied to the electrodes, as shown by figure 4.*

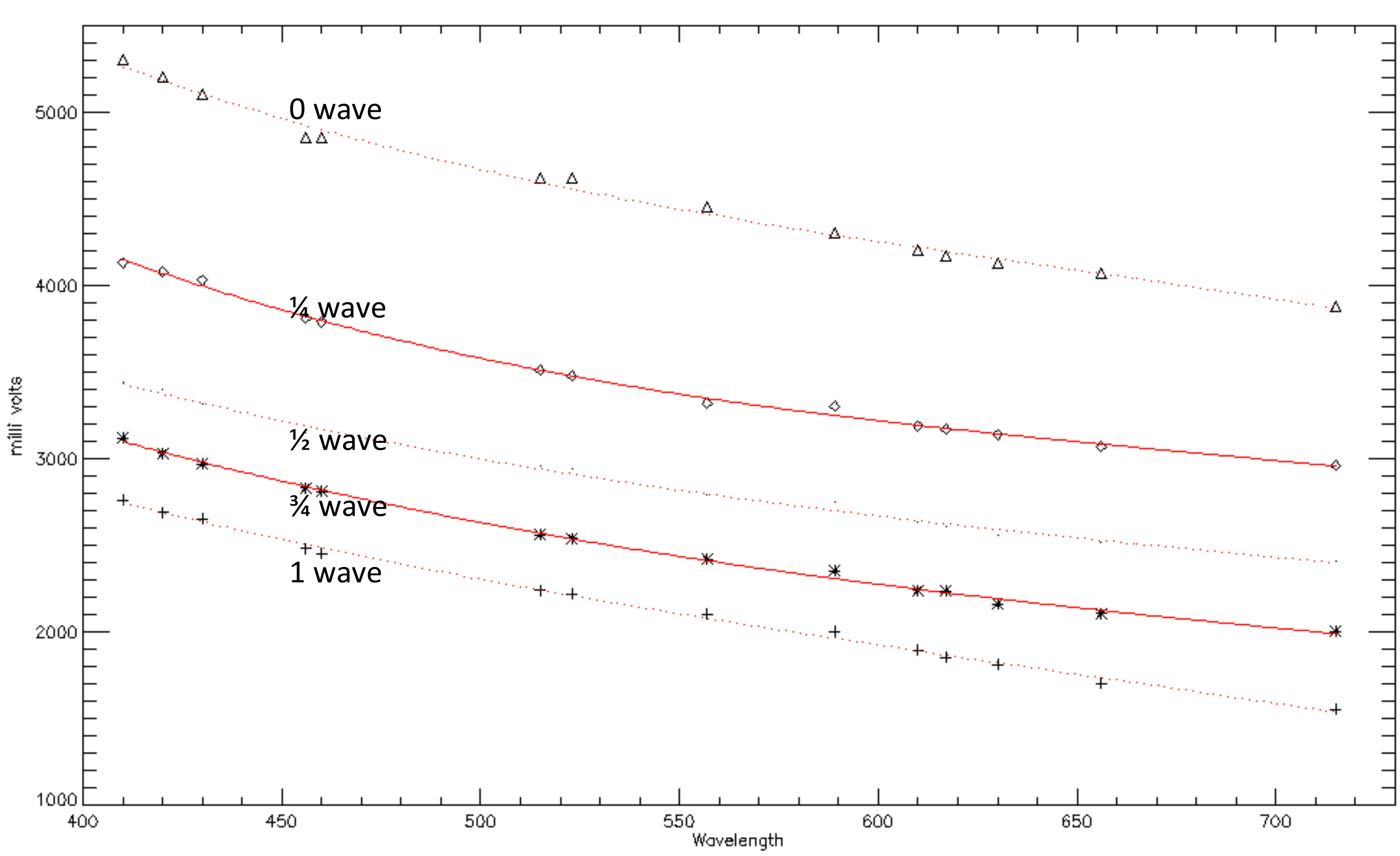

*Figure 4 : Electric voltage applied to the LCVR for 0, 1/4, 1/2, 3/4, 1 wave retardance as a function of wavelength. In the present case, the modulation was between 0 and half wave. The voltage is adjusted to the wavelength of the observed line.*

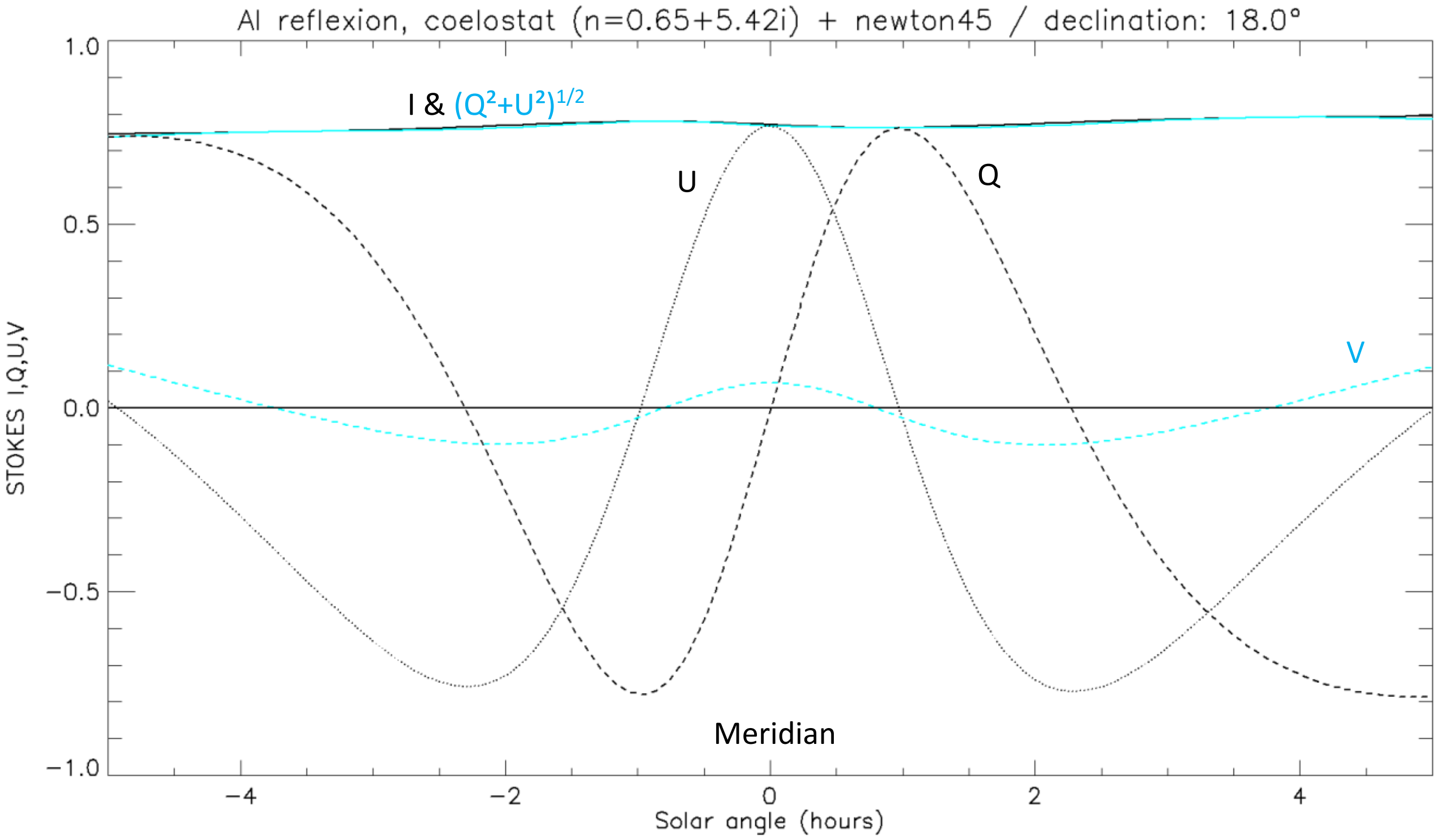

*Figure 5 : I, Q, U, V at the focus of the telescope for an incident light with Q = V = 0. $(Q^2+U^2)^{1/2}$ remains almost constant but a crosstalk between U and Q could occur during the diurnal motion. Theoretical calculation for 18° declination and 460 nm wavelength.*

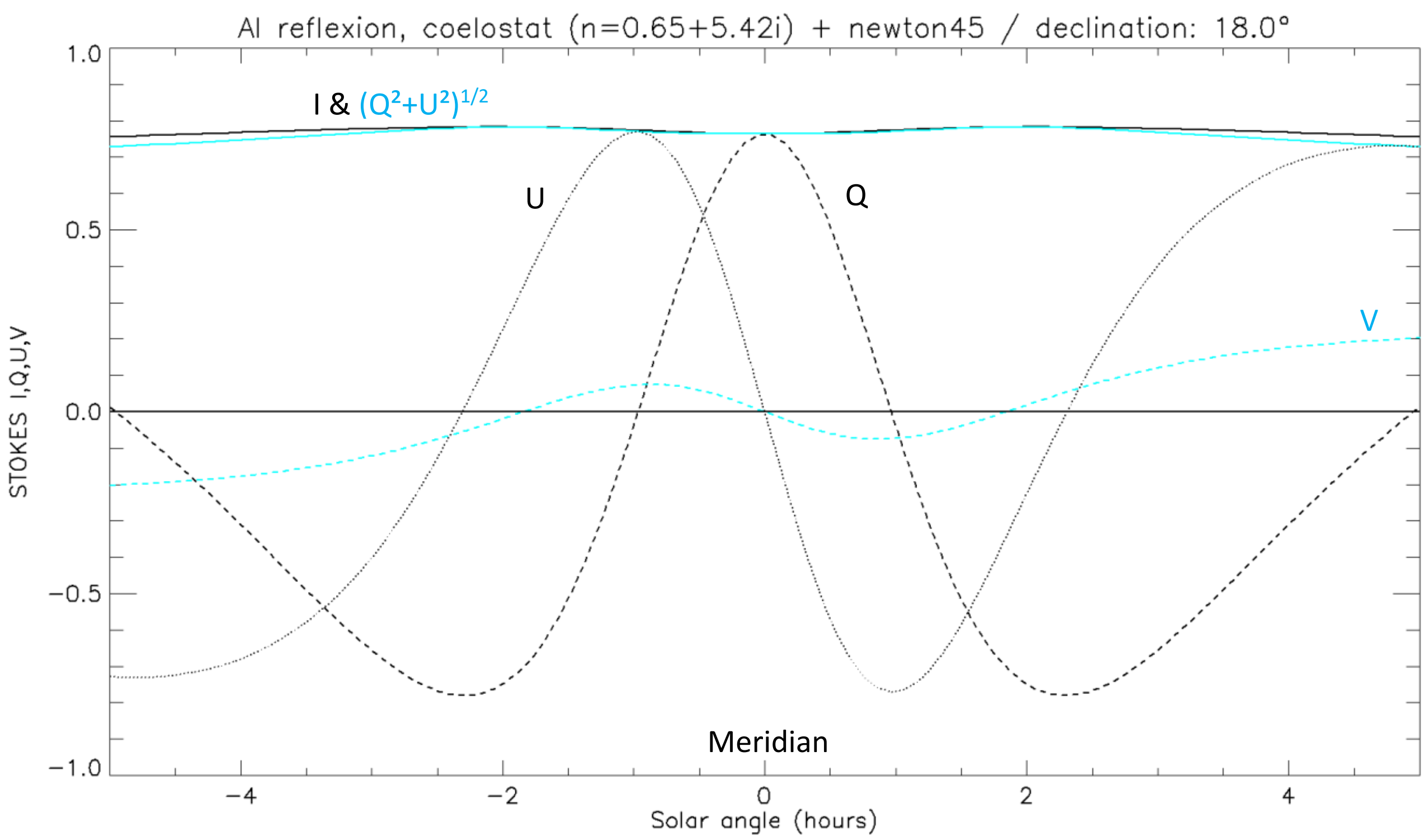

*Figure 6 : I, Q, U, V at the focus of the telescope for an incident light with U = V = 0. $(Q^2+U^2)^{1/2}$ remains almost constant but a crosstalk between U and Q could occur during the diurnal motion. Theoretical calculation for 18° declination and 460 nm wavelength.*

## III – CaI 4427 A AT 5” OF THE LIMB

Two runs were performed for CaI 4227 A, the first one with the slit at 5” of the limb and the second one at 20”. At 5”, the distance to the limb of the different parts of the slit is variable, because of the curvature of the limb: when the central part is at 5”, both ends are very close to the limb or at the limb. For that reason, we present the results of the central part and ends separately. On the contrary, at 20”, the gradient of polarization along the slit becomes negligible and we present the result after full integration along it.

### III - 1 - At 5” of the limb (μ = 0.1)

Figure 7 shows the accumulation of 36 observations (36 couples I+Q, I-Q) of the central part (360 pixels) of the slit with 20 s exposure time. After integration along the slit (figure 8), the noise is strongly reduced by a factor 15 to 20 and we estimate the SNR at 13000 in the continuum and 3000 in the line core. Results are comparable to those of Bianda *et al* (1999).

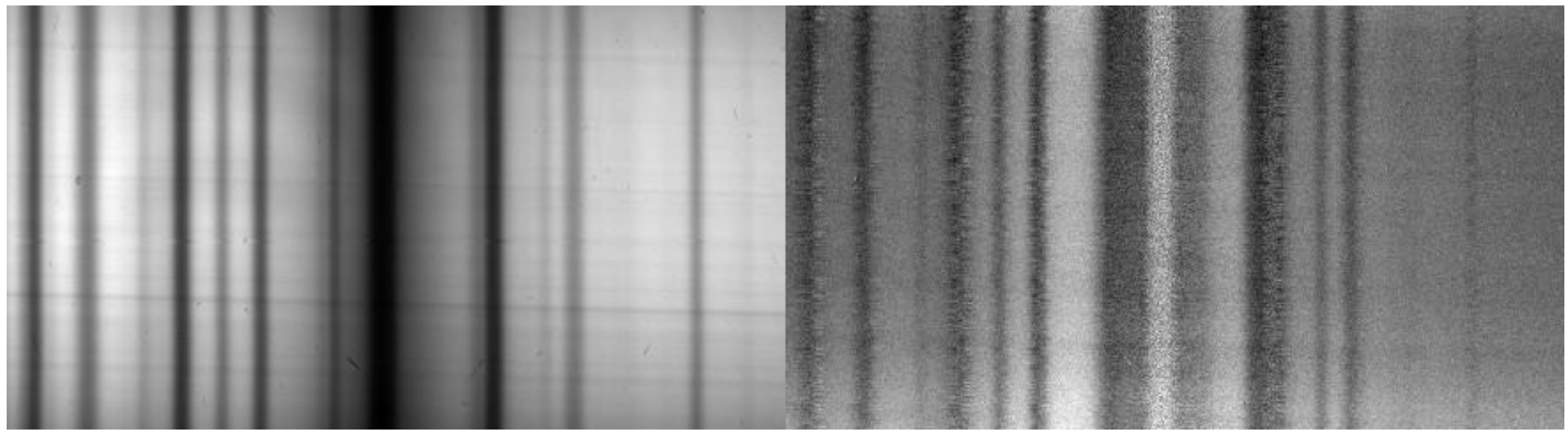

*Figure 7 : intensity and Q/I for CaI 4227 A at 5” of the limb. Wavelength in abscissa, and the slit direction in ordinates. 7 May 2008, 10:48 to 11:13 UT.*

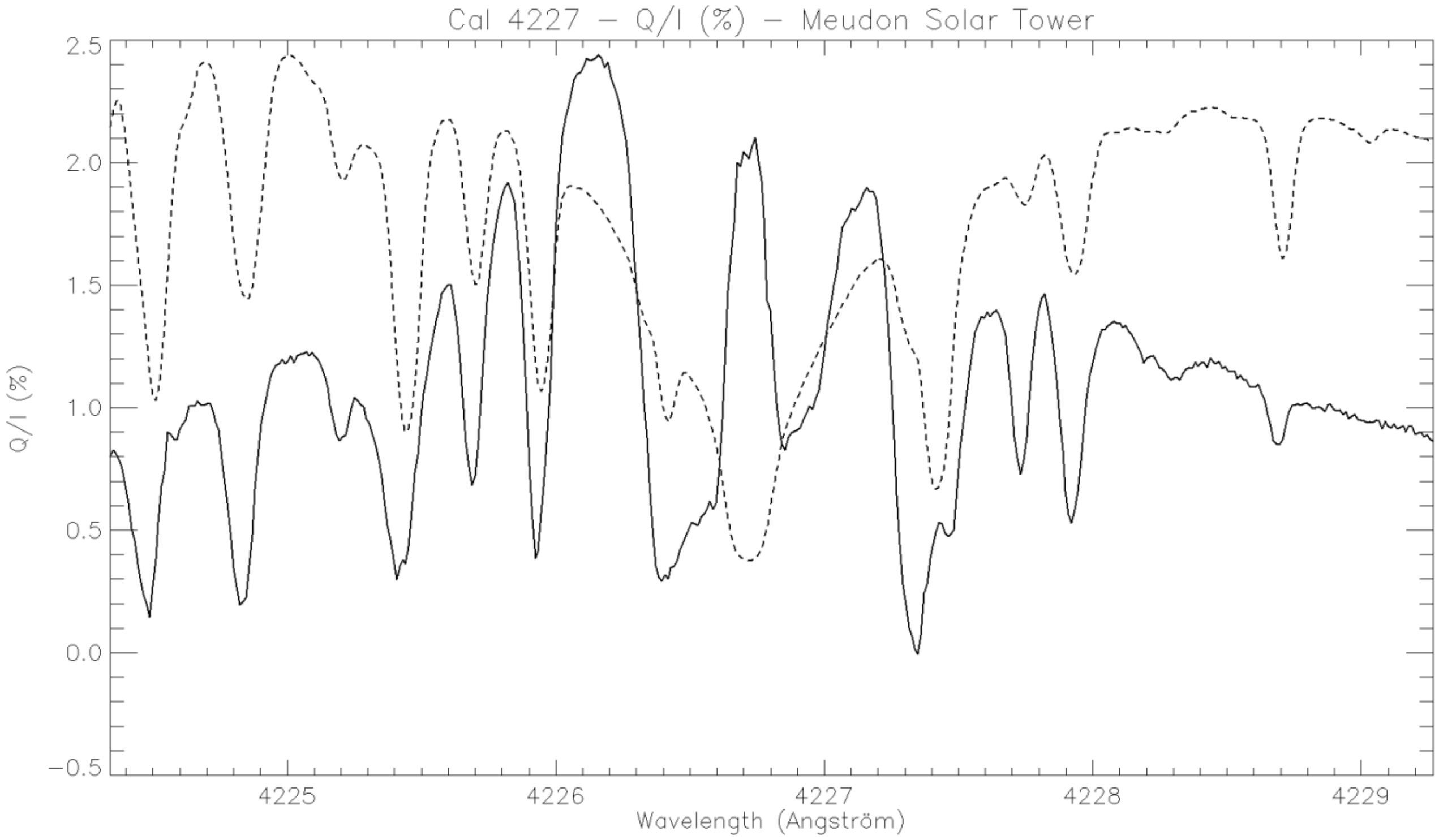

*Figure 8 : intensity (dashed line) and Q/I (solid line) in % for CaI 4227 A at 5” of the limb after integration along the slit. 7 May 2008, 10:48 to 11:13 UT.*

### III - 2 - Close to the limb (μ < 0.1)

Figure 9 shows the accumulation of 36 observations of one end (20 pixels) of the slit. After integration along 20 pixels (figure 10), the noise is a little bit reduced and we see that the polarization Q/I reaches 3% in the line core. It could even be higher, without crosstalk effects of the coelostat.

*Figure 9 : intensity and Q/I for CaI 4227 A close to the limb. Wavelength in abscissa, and the slit direction in ordinates. 7 May 2008, 10:48 UT to 11:13 UT.*

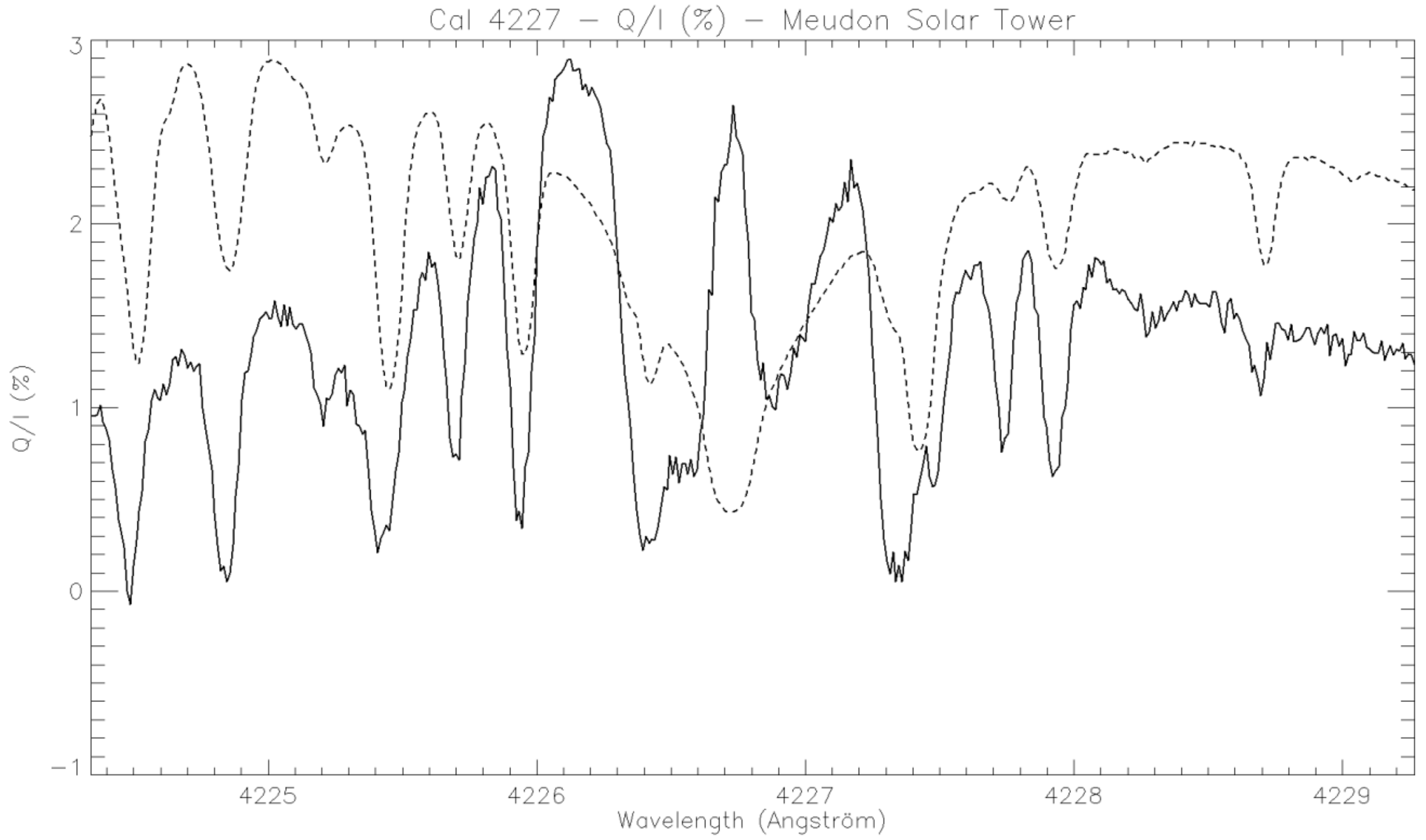

*Figure 10 : intensity (dashed line) and Q/I (solid line) in % for CaI 4227 A close to the limb. 7 May 2008, 10:48 UT to 11:13 UT.*

### III - 3 - At 20" of the limb (μ = 0.2)

Figure 11 shows the accumulation of 100 observations (100 couples I+Q, I-Q) of the slit with 15 s exposure time. After full integration along the slit (figure 12), the noise is strongly reduced by a factor 20 and we estimate the SNR at 16000 in the continuum and 6000 in the line core.

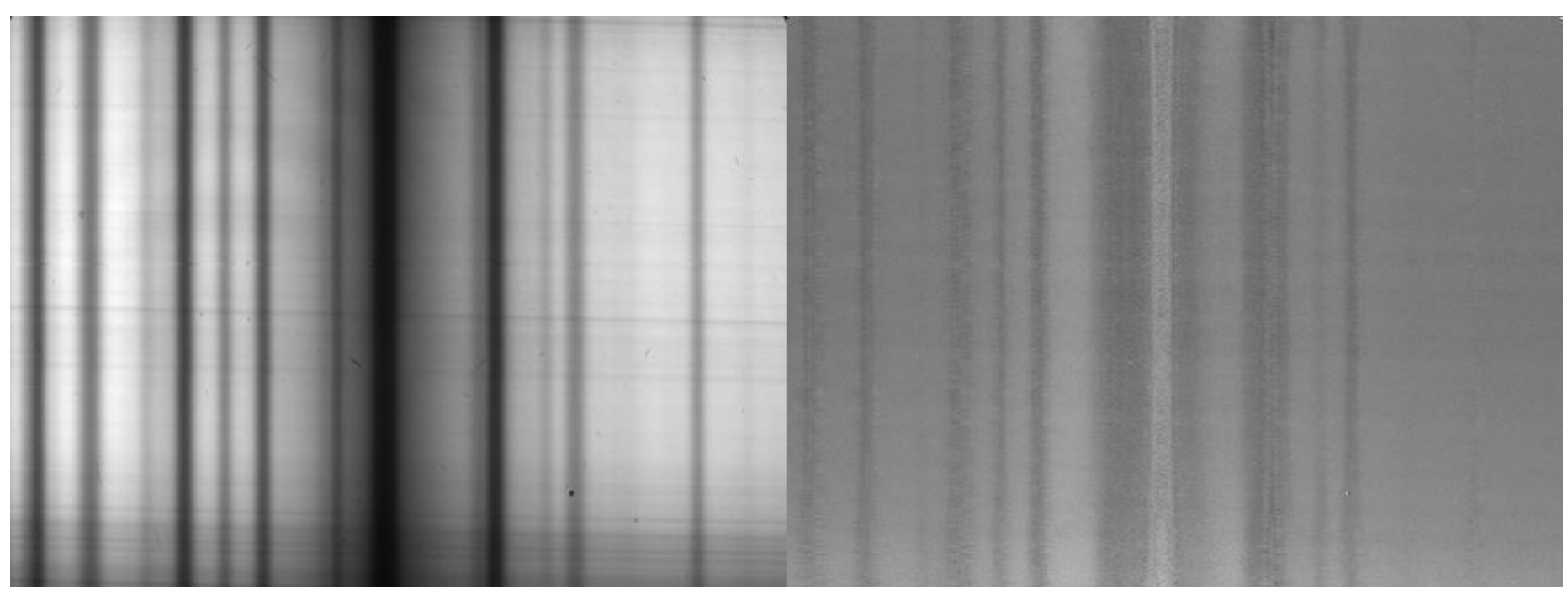

*Figure 11 : intensity and Q/I for CaI 4227 A at 20" of the limb. Wavelength in abscissa, and the slit direction in ordinates. 7 May 2008, 09:43 UT to 10:37 UT.*

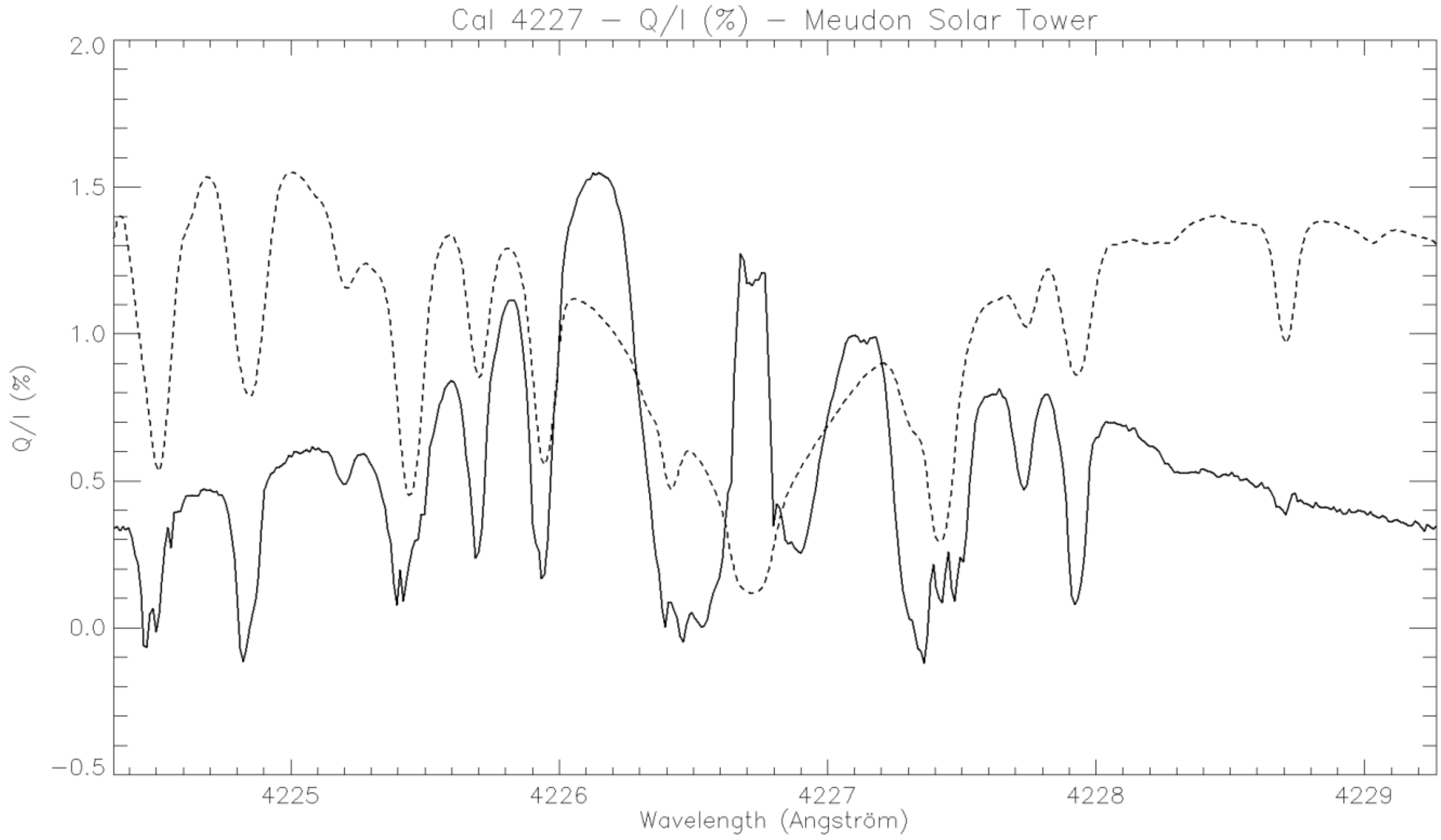

*Figure 12 : intensity (dashed line) and Q/I (solid line) in % for CaI 4227 A at 20" of the limb after integration along the slit. 7 May 2008, 09:43 UT to 10:37 UT.*

## IV – BaII 4554 A AT 20" OF THE LIMB (μ = 0.2)

Figure 13 shows the accumulation of 100 observations (100 couples I+Q, I-Q) of the slit with 4 s exposure time. After full integration it (figure 14), the noise is strongly reduced and we estimate the SNR at 39000 in the continuum and 19000 in the line core. Faurobert *et al* (2009) studied this line from a theoretical point of view.

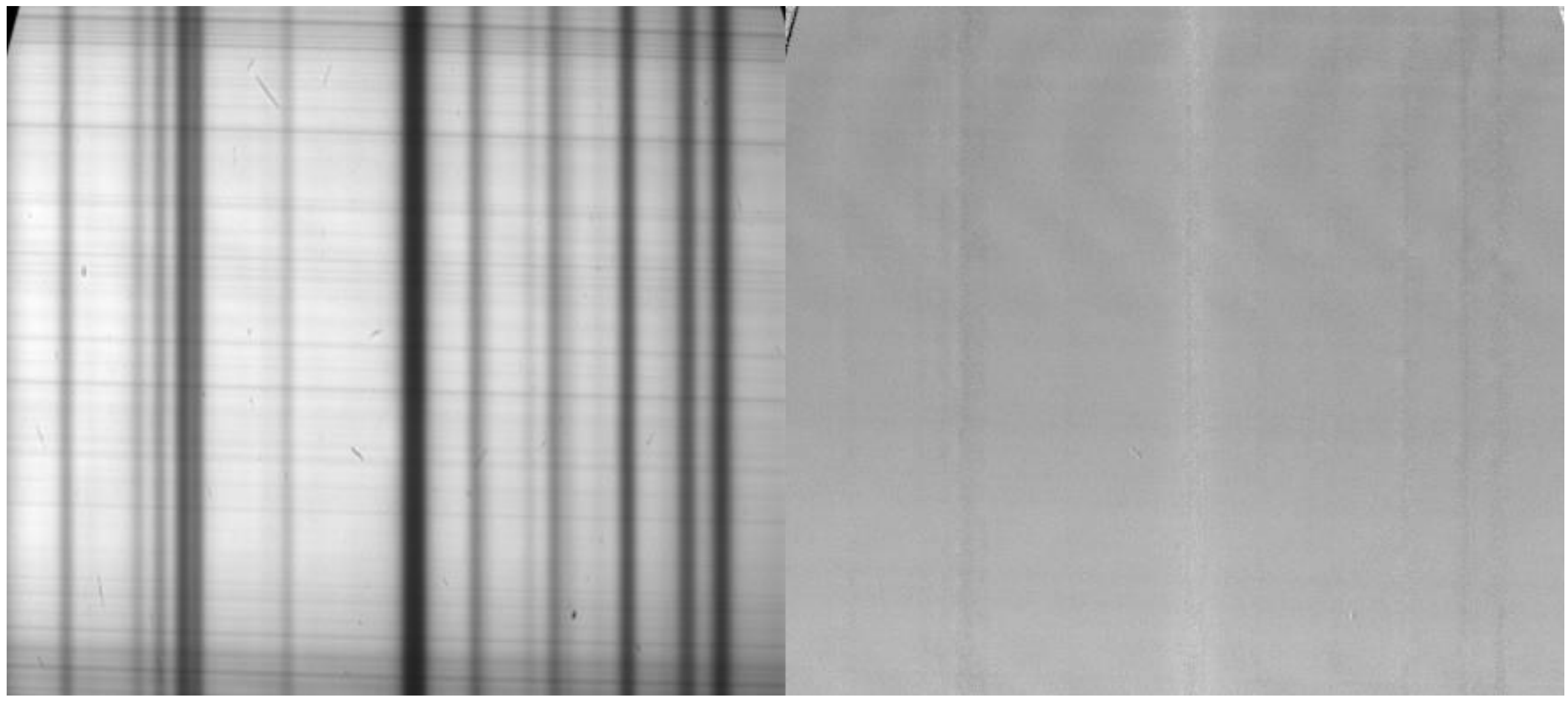

*Figure 13 : intensity and Q/I for BaII 4554 A at 20" of the limb. Wavelength in abscissa, and the slit direction in ordinates. 6 May 2008, 13:00 UT to 13:18 UT.*

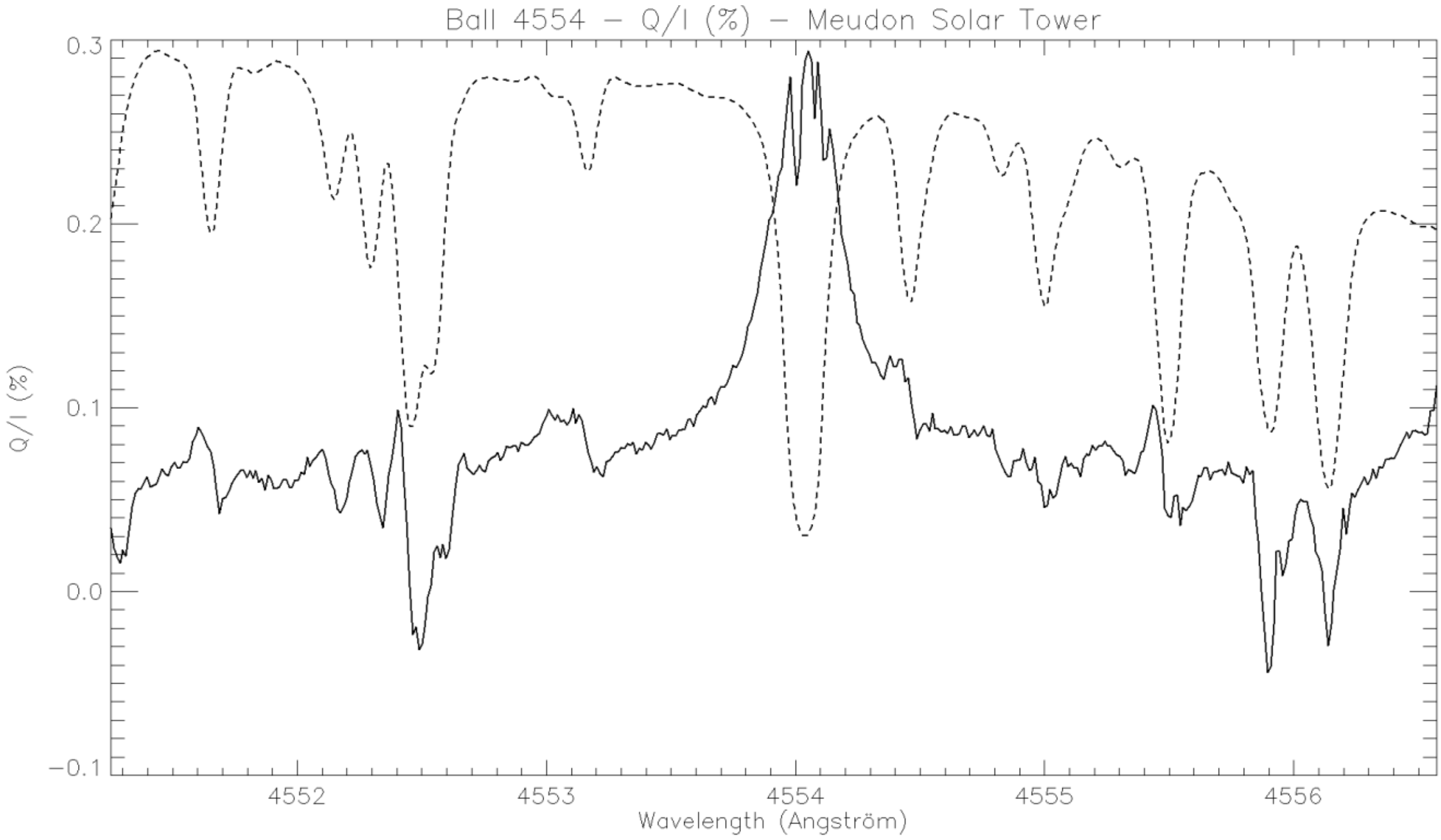

*Figure 14 : intensity (dashed line) and Q/I (solid line) in % for BaII 4554 A at 20" of the limb after integration along the slit. 6 May 2008, 13:00 UT to 13:18 UT. The hyperfine structure of the line core (isotopes) is visible.*

**V – SrI 4607 A AT 20" OF THE LIMB (μ = 0.2)**

Figure 15 shows the accumulation of 100 observations (100 couples I+Q, I-Q) of the slit with 4 s exposure time. After full integration along the pixels of the slit (figure 16), the noise is strongly reduced and we estimate the SNR at 29000 in the continuum and 24000 in the line core. Derouich *et al* (2006) showed how to use SrI 4607 to probe the turbulent magnetic field with depth.

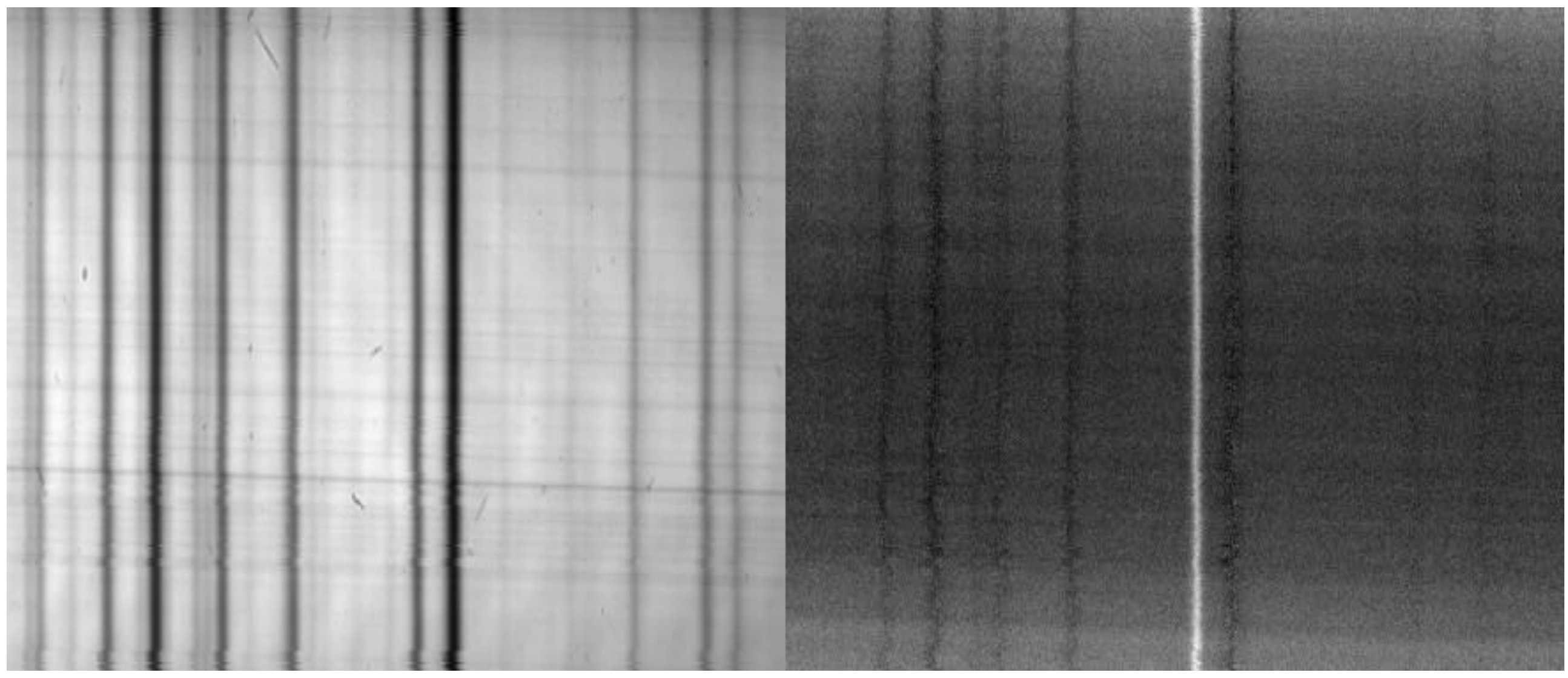

*Figure 15 : intensity and Q/I for SrI 4607 A at 20" of the limb. Wavelength in abscissa, and the slit direction in ordinates. 6 May 2008, 08:23 UT to 08:41 UT.*

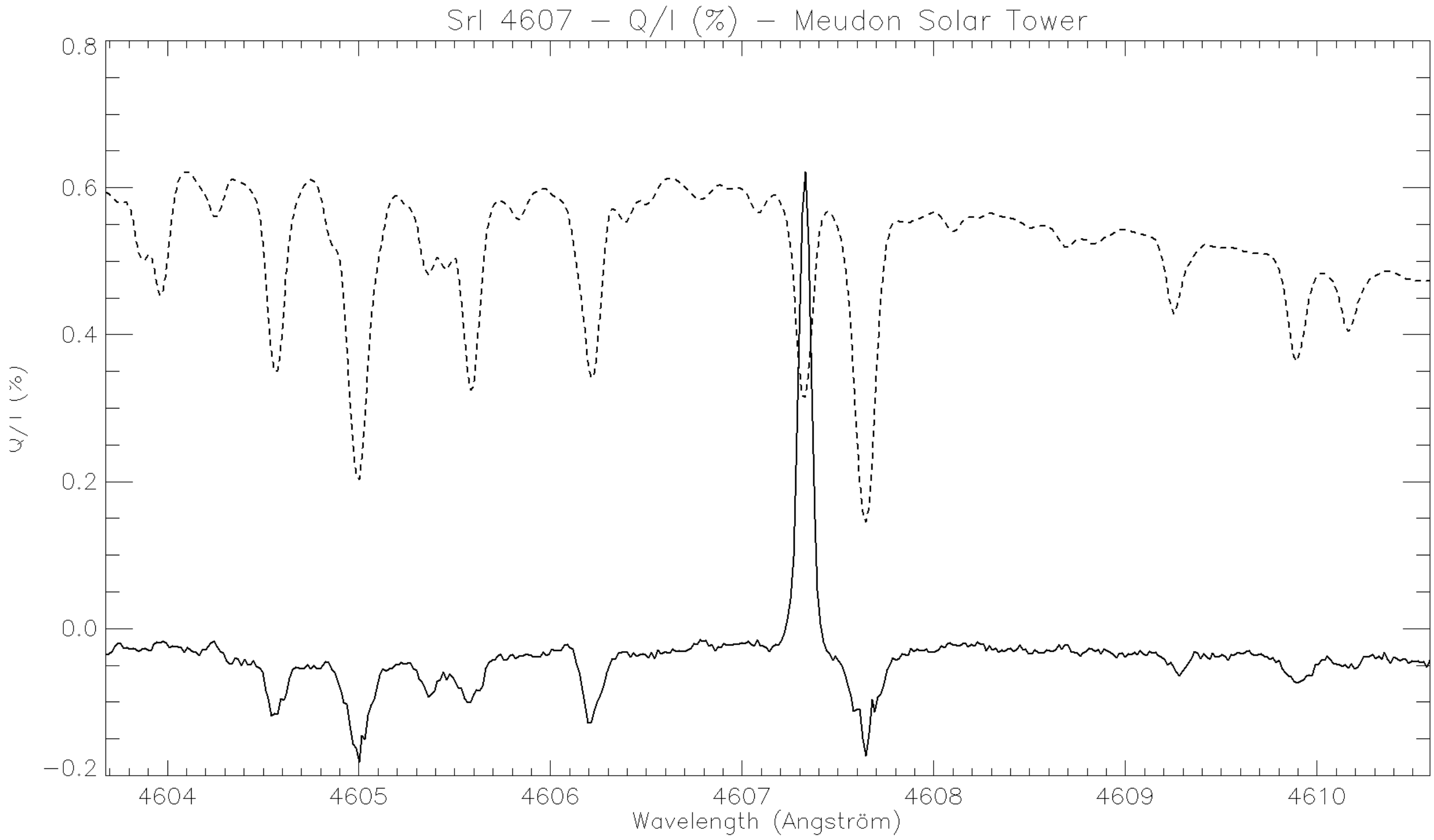

*Figure 16 : intensity (dashed line) and Q/I (solid line) in % for SrI 4607 A at 20" of the limb after integration along the slit. 6 May 2008, 08:23 UT to 08:41 UT.*

## VI – $C_2$ & MgH MOLECULES AT 5140 A & 20" OF THE LIMB (μ = 0.2)

Figure 17 shows the accumulation of 100 observations (100 couples I+Q, I-Q) of the slit with 0.6 s exposure time. After full integration along all pixels of the slit (figure 18), the noise is reduced and we estimate the SNR at 50000 in the continuum and 29000 in the line cores of deep lines (Fe, Ni). Molecular lines are indeed very faint in intensity but not in polarization.

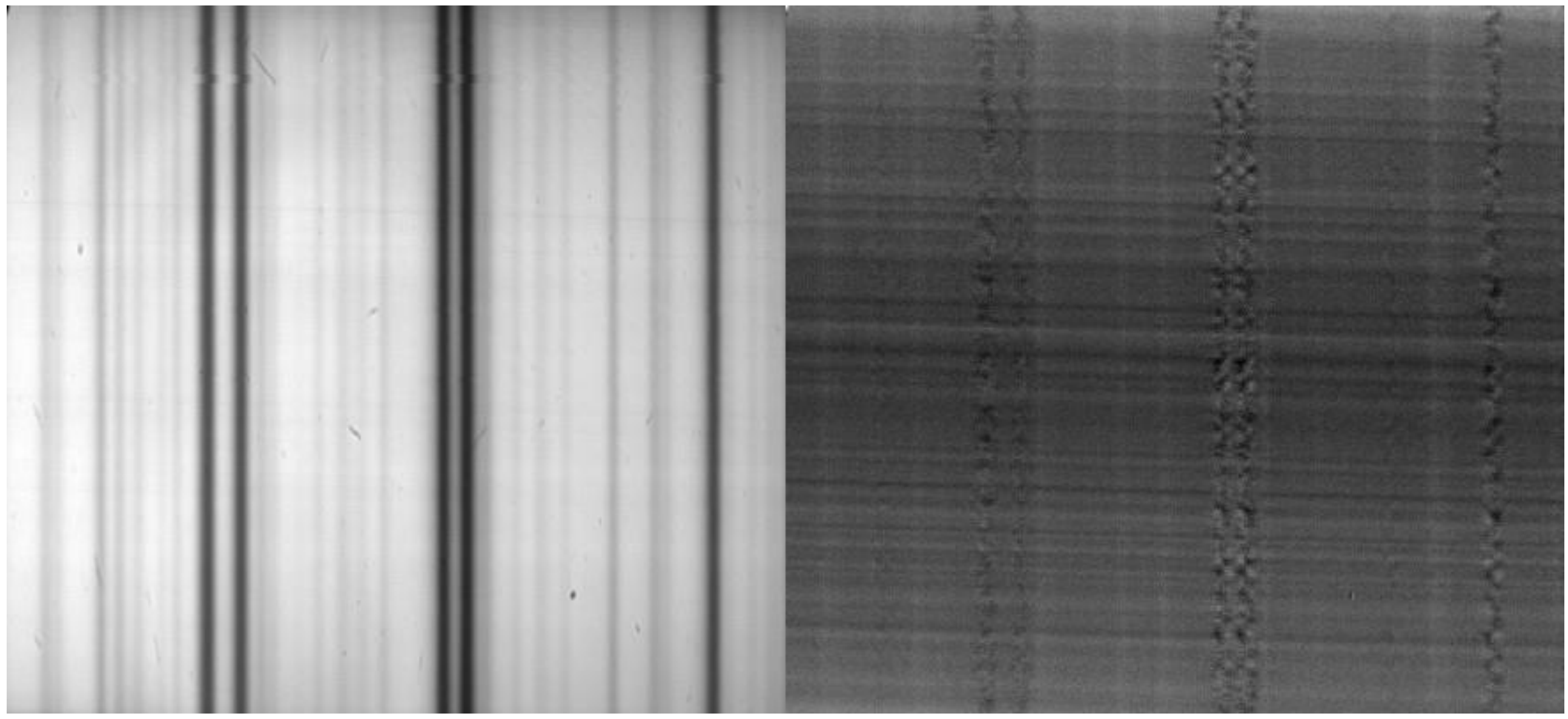

*Figure 17 : intensity and Q/I for $C_2$ & MgH molecules around 5140 A and at 20" of the limb. Wavelength in abscissa, and the slit direction in ordinates. 13 May 2008, 10:30 UT to 10:37 UT.*

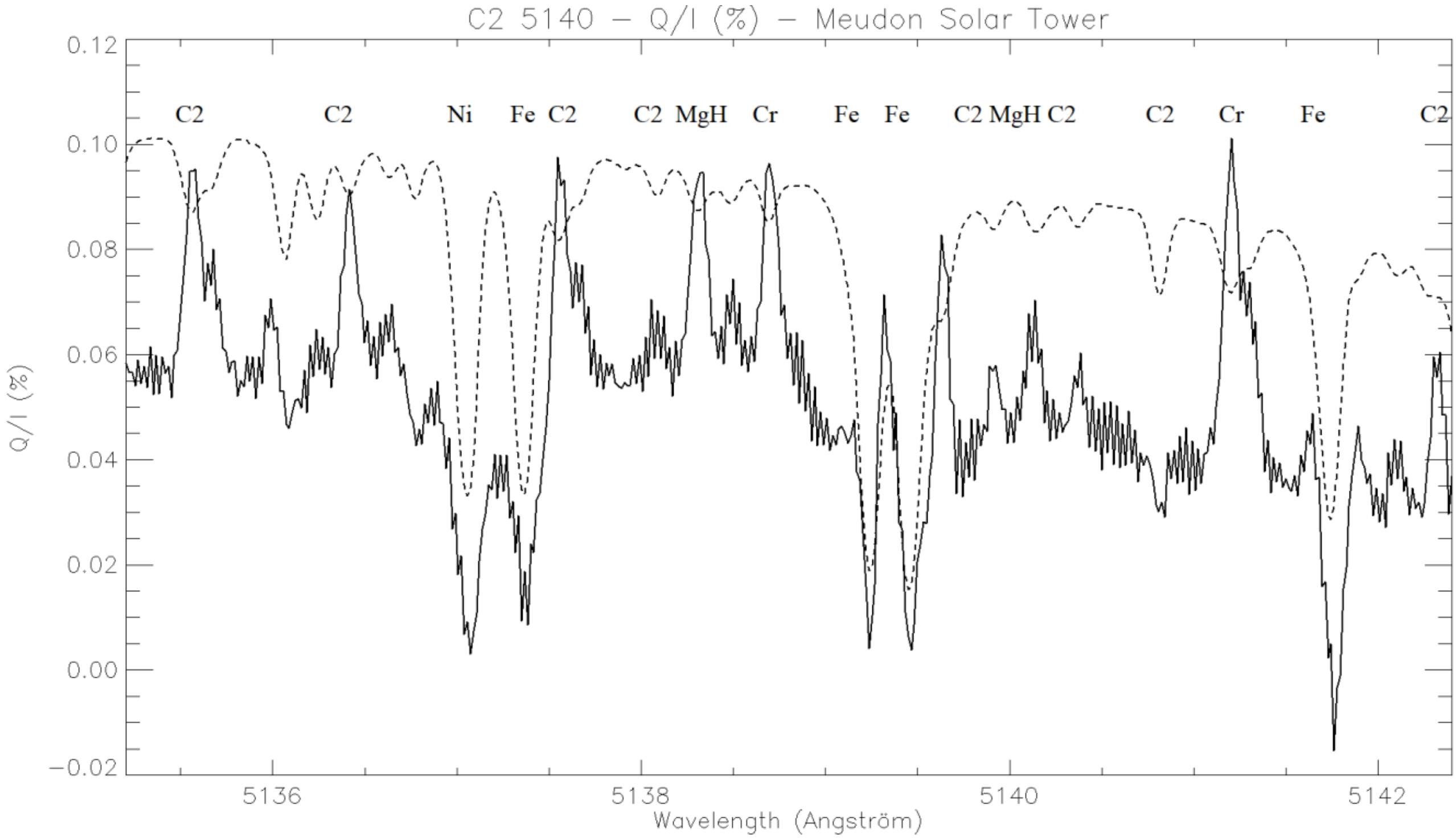

*Figure 18 : intensity (dashed line) and Q/I (solid line) in % for $C_2$ & MgH molecules around 5140 A at 20" of the limb after integration along the slit. 13 May 2008, 10:30 UT to 10:37 UT. Strong lines of Ni and Fe are depolarized. $C_2$ & MgH are very faint lines but exhibit polarization.*

## VII – NdII AT 5250 A & 20" OF THE LIMB (μ = 0.2)

Figure 19 shows the accumulation of 100 observations (100 couples I+Q, I-Q) of the slit with 0.5 s exposure time. After full integration along the slit (figure 20 & 21), the noise is strongly reduced and we estimate the SNR at 50000 in the continuum and 30000 in the line cores of deep lines (Fe, Ni). Two sequences are available (13 May 2008, 08:01 UT to 08:07 UT & 08:15 UT to 08:30 UT); the second one is two times longer, with 200 couples (I+Q, I-Q).

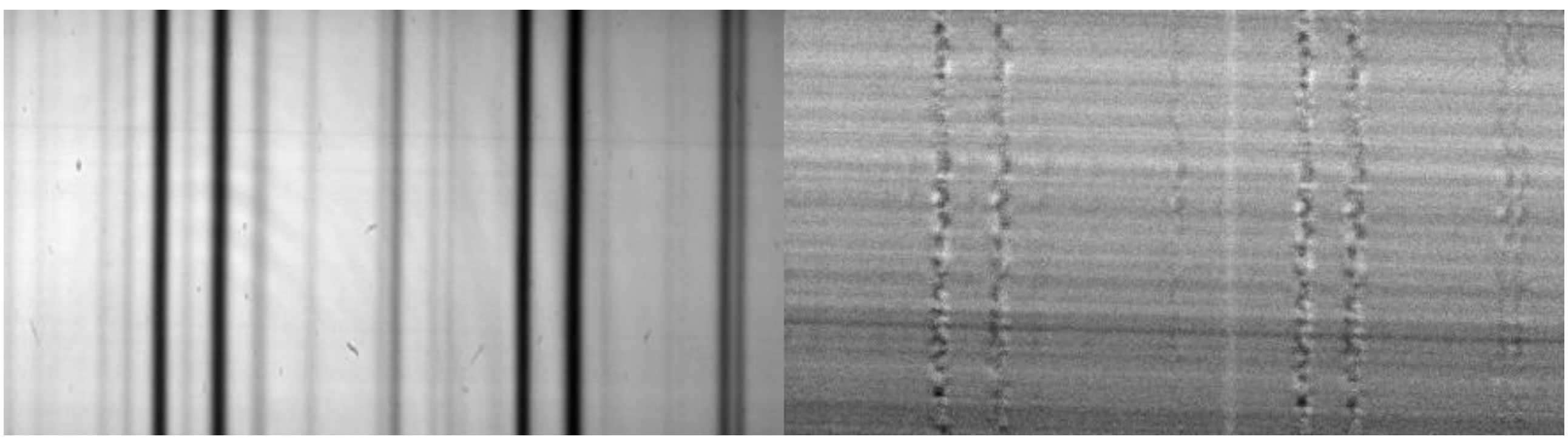

*Figure 19 : intensity and Q/I for NdII at 5250 A and at 20" of the limb. Wavelength in abscissa, and the slit direction in ordinates. 13 May 2008, 08:15 UT to 08:30 UT. The polarization of NdII is well visible at right, but this is a faint line in intensity.*

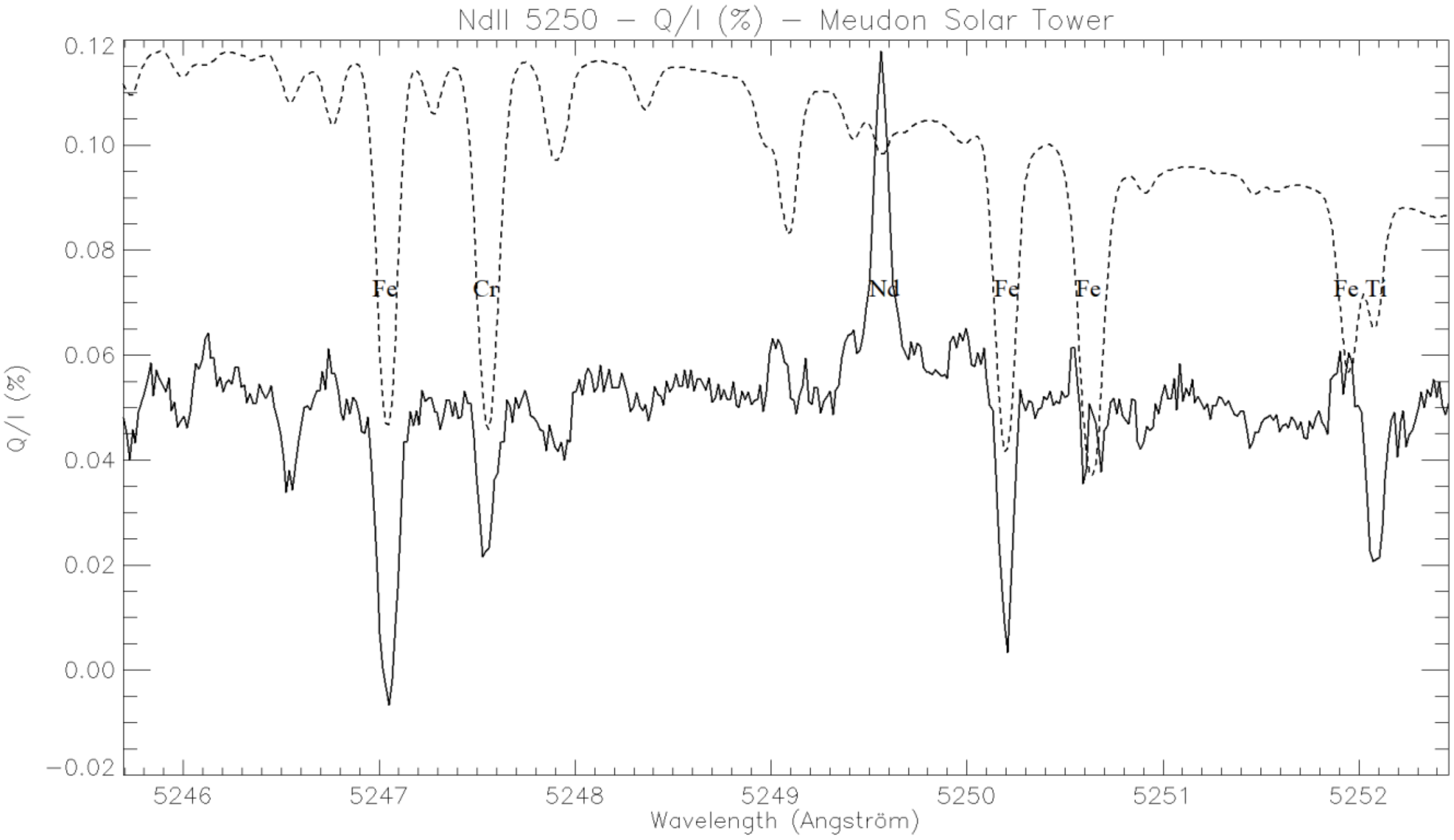

*Figure 20 : intensity (dashed line) and Q/I (solid line) in % for NdII 5250 A at 20" of the limb after integration along the slit. 13 May 2008, 08:01 UT to 08:07 UT. Strong lines of Cr and Fe are depolarized. NdII is a very faint line but exhibit polarization.*

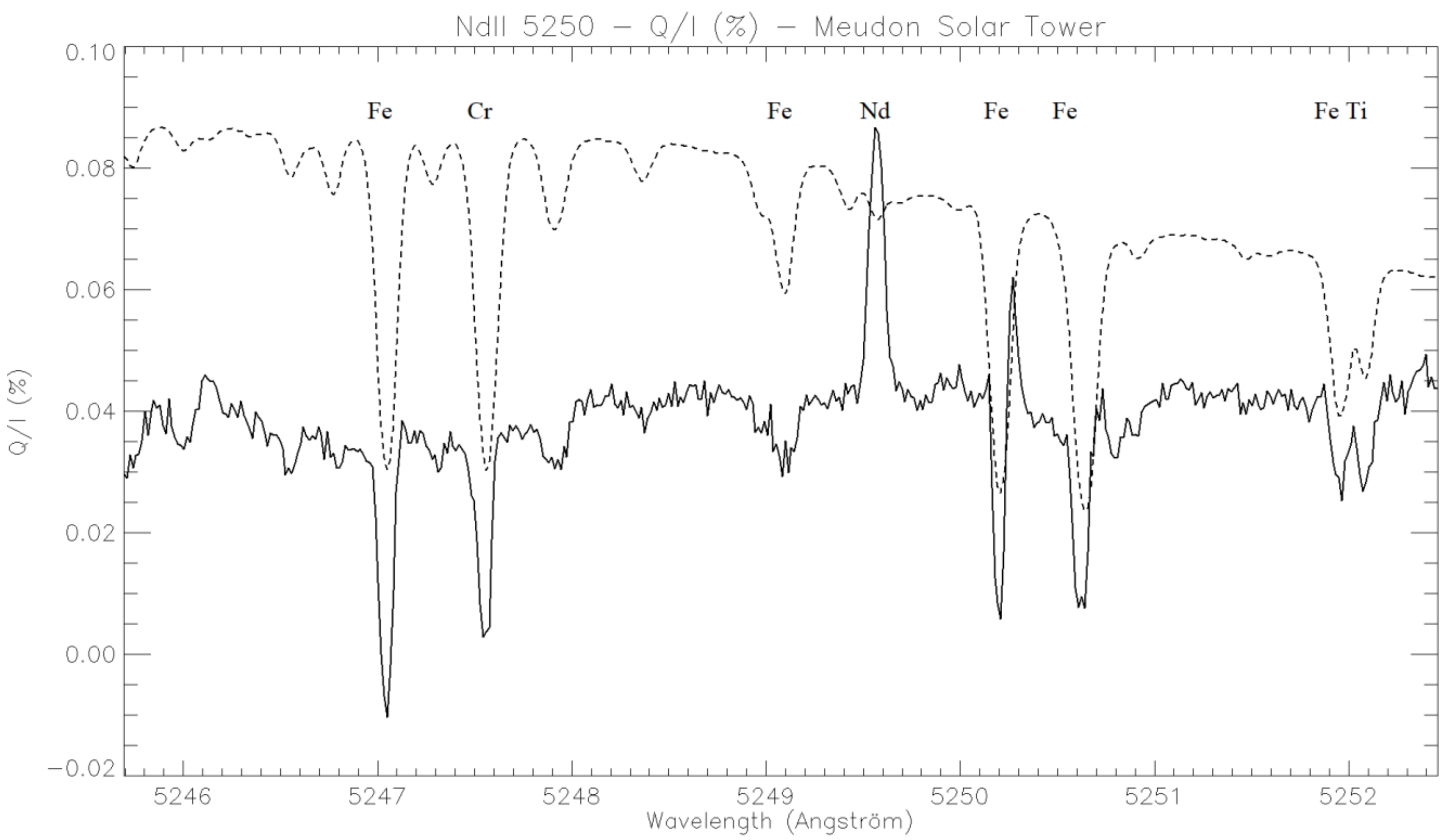

*Figure 21 : intensity (dashed line) and Q/I (solid line) in % for NdII 5250 A at 20" of the limb after integration along the slit. 13 May 2008, 08:15 UT to 08:30 UT. Strong lines of Cr and Fe are depolarized. NdII is a very faint line but exhibit polarization.*

## VIII – THE DATASET

The present data are available on line in scientific FITS format with a description at :

https://entrepot.recherche.data.gouv.fr/dataset.xhtml?persistentId=doi:10.57745/YK18JE

## IX - CONCLUSION

The present data were obtained with a liquid crystal polarimeter and a classical CCD camera with mechanical shutter running at 1 Hz cadence. Stokes combination I+Q and I-Q were obtained sequentially in the form of 2D spectral images (λ, x); then tens or hundreds of such frames were got in sequence and accumulated; then pixels along the slit were added, in order to achieve a final SNR much better than 10000. The observed lines are of interest for measurements of weak unresolved magnetic fields and variations with depth, through the Hanle depolarization.

## X - REFERENCES

Bianda, M., Stenflo, J.O., Solanki, S.K., 1999, "Hanle effect observations with the CaI 4227 A line", *Astron. Astrophys*., 350, 1060

Derouich, M., Bommier, V., Malherbe, J.M., Landi Degl'Innocenti, E., 2006, "Second solar spectrum of The SrI 4607 A line: depth probing of the turbulent magnetic field strength in a quiet region", *Astron. Astrophys*, 457, 1047

Faurobert, M., Derouich, M., Bommier, V., Arnaud, J., 2009, "Hanle effect in the solar BaII D2 line: a diagnostic tool for chromospheric weak magnetic fields", *Astron. Astrophys*., 493, 201

Malherbe, J.-M., Roudier, Th., Moity, J., Mein, P., Arnaud, J., Muller, R., 2007, "Spectropolarimetry with liquid crystals", *Mem. S. A. It*., 78, 203.

Stenflo, J.O., 1982, "The Hanle effect and the diagnostics of turbulent magnetic fields in the solar atmodphere", *Solar Phys*., 80, 209

Stenflo, J., Keller, C., 1997, "The second solar spectrum. A new window for diagnostics of the Sun", *Astron. Astrophys*., 321, 927

Stenflo, J, 2004, "a new world of scattering physics seen by high precision imaging polarimetry", *Reviews in Modern Astronomy*, 17, 269